\documentclass[10pt, a4paper]{article}

\usepackage[final]{lrec-coling2024} 

\usepackage{tikz}
\usepackage{dblfloatfix}
\usepackage{pgfplots}
\usepackage{caption}
\usepackage{booktabs}
\usepackage{multirow}
\usepackage{graphicx}
\usepackage{subfig}

\usepackage{pifont}
\newcommand{\cmark}{\ding{51}}%
\newcommand{\xmark}{\ding{55}}%

\title{\vspace*{.5\baselineskip} \textbf{BEIR-PL: Zero Shot Information Retrieval Benchmark
for the Polish Language}}

\name{Konrad Wojtasik, Vadim Shishkin, Kacper Wołowiec, \\ {\bf \large Arkadiusz Janz, Maciej Piasecki}}

\address{Wrocław University of Science and Technology \\
         konrad.wojtasik@pwr.edu.pl \\
         }

\abstract{
The BEIR dataset is a large, heterogeneous benchmark for Information Retrieval (IR), garnering considerable attention within the research community. However, BEIR and analogous datasets are predominantly restricted to English language. Our objective is to establish extensive large-scale resources for IR in the Polish language, thereby advancing the research in this NLP area. In this work, inspired by mMARCO and Mr.~TyDi datasets, we translated all accessible open IR datasets into Polish, and we introduced the BEIR-PL benchmark -- a new benchmark which comprises 13 datasets, facilitating further development, training and evaluation of modern Polish language models for IR tasks. We executed an evaluation and comparison of numerous IR models on the newly introduced BEIR-PL benchmark. Furthermore, we publish pre-trained open IR models for Polish language, marking a pioneering development in this field. 
The BEIR-PL is included in MTEB Benchmark and also available with trained models at URL {\url{https://huggingface.co/clarin-knext}}.
 \\ \newline  }

\begin{document}

\maketitleabstract

\section{Introduction}

Modern natural language processing (NLP) applications often require support from efficient information retrieval (IR) processes, e.g.\ in order to efficiently acquire and accurately pre-filter texts. An IR component is necessary in the case of many NLP tasks such as Question Answering, Entity Linking, or Abstractive Summarization. Recently, classic IR models based on lexical matching are typically combined with neural retrievers utilizing large pre-trained language models. The neural language models based on the transformer architecture facilitate solving NLP problems in a multilingual setting due to their cross-lingual alignment originating from pre-training on parallel corpora. 
Such models have achieved promising results in a zero-shot setting, in which the model is trained only on the source language data only and evaluated on the target. 
Thus, there is a great need to create cross-lingual evaluation data and benchmarks similar to the monolingual BEIR \cite{thakur-2021-BEIR} for many languages. Although existing multilingual evaluation benchmarks try to include as many languages as possible, the Polish language has been much less prominent in IR studies focused on neural models due to the limited availability of Polish IR datasets.
The existing studies on dense IR tasks suggest that dense retrieval models outperform lexical matchers, such as BM25 algorithm \cite{DBLP:journals/corr/abs-2004-04906}. However, measuring the performance gap between lexical retrievers and dense retrieval models is very instructive, as the lexical matchers typically require much fewer computational resources. Moreover, lexical matchers are still a tough baseline for neural retrievers in specific domains.

\begin{figure*}[hbtp]
\centering
\includegraphics[scale=0.23]{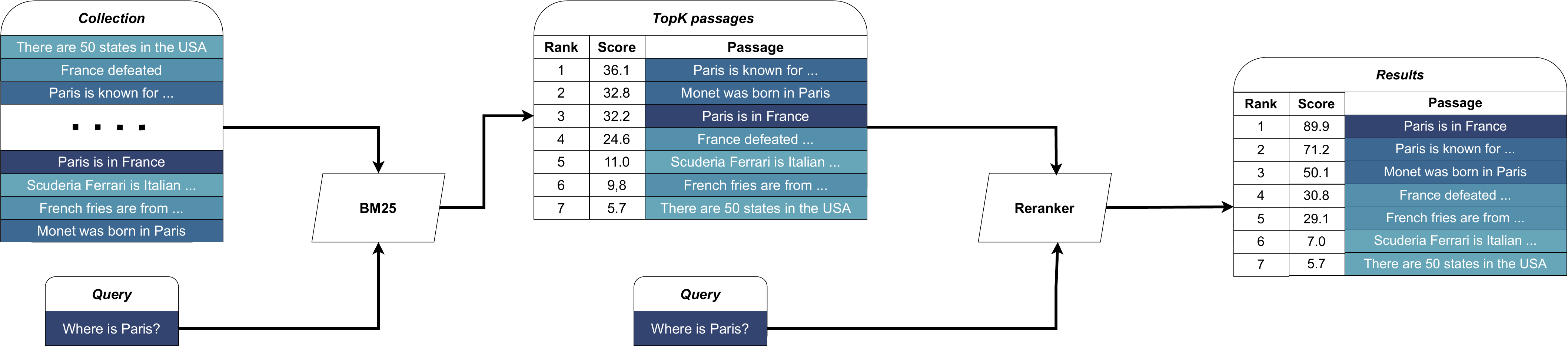}
\caption{\label{fig:retriev-rerank} In retrieval with re-ranking setting, in the first stage, top@k most relevant documents are retrieved by the fast but less accurate model. In our case, it was BM25. Afterward, the documents are re-ranked by a more powerful and more accurate model.}
\label{fig:x cubed graph}
\end{figure*}

Our main goal was to create a large scale benchmark dataset for IR in the Polish language, which is especially aimed at zero-shot approaches. Additionally, we wanted to train and evaluate existing IR models known from literature, but have not been so thoroughly analysed on Polish datasets, yet, to determine their performance and establish a baseline for future research.
Our contributions are as follows.
\begin{itemize}
    \item For the sake of comparison and compatibility, 
we translated the original BEIR benchmark datasets to the Polish language,
a less-resourced language in IR,
e.g., \ Polish was neither covered by the original multilingual MS~MARCO dataset \cite{DBLP:journals/corr/NguyenRSGTMD16}, nor by Mr.~TyDi benchmark \cite{DBLP:journals/corr/abs-2108-08787}, that 
had been a strongly limiting factor for developing dense retrieval models for the Polish language. 
    \item We fine-tuned five deep neural models for the re-ranking task using different model architectures and sizes, that are currently present in the literature.
    \item We fine-tuned an unsupervised dense bi-encoder as a retriever on the Inverse Cloze Task (ICT) and compared its performance with an available multilingual sentence embedding model, as well as with lexical BM25 retrieval.
    \item We demonstrated that both the BEIR-PL and original BEIR benchmarks are of heterogeneous nature, and that to accurately compare model performance, it is necessary to closely examine the results of individual datasets rather than relying solely on overall averages across the entire dataset.
    \item We tested trained models on PolEval 2022 Passage Retrieval competition datasets, which cover three different domains and provides additional testing for trained models.
\end{itemize}

\section{Related work}
We built upon the idea of the BEIR benchmark dataset \cite{thakur-2021-BEIR}, as it is precisely focused on the zero-shot evaluation of modern IR systems. Neural IR systems are trained on large datasets such as MS~MARCO \cite{DBLP:journals/corr/abs-2108-13897}, or synthetic datasets derived from large pre-trained generative language models \cite{10.1145/3477495.3531863}.  MS~MARCO has been translated into many different languages \cite{DBLP:journals/corr/abs-2108-13897}, but not to Polish, yet. 
Moreover, even other extensive multilingual benchmarks for IR such as Mr.~TyDi \cite{DBLP:journals/corr/abs-2108-08787} -- 
covering many topologically diverse languages -- do not include Polish data and includes so far only one Slavic language, namely Russian. There is the MTEB Benchmark \cite{muennighoff2022mteb}, which is a benchmark dedicated semantic embeddings, where BEIR-PL is included as part of the Polish retrieval datasets. 
The two most commonly used and recent benchmarks for the Polish language technology, namely KLEJ \cite{DBLP:journals/corr/abs-2005-00630} and Lepiszcze \cite{lepiszcze} contain evaluation data for many distinct NLP tasks, but none of them is directly related to IR tasks. The recently published large resource for IR in Polish language, MaupQA \cite{rybak-2023-maupqa}, provides nearly 400k question-passage pairs collected from different resources. Moreover, there was Passage Retrieval task in PolEval 2022 competition \cite{KNOWCON20235627}, driving research in open domain IR for Polish language pushing researchers to develop effective solutions \cite{KNOWCON20239253_pokrywka,KNOWCON20238119_kozlowski,KNOWCON20233900_wojtasik}. 

\subsection{Passage Retrieval}

The task of textual IR is to search for and return documents (i.e.\ any indexed text objects) that are relevant to a user query from a collection. Collections may consist of millions of documents, which makes the task computationally intensive. Moreover, documents and queries mostly are of significantly different lengths, the language used throughout the documents may vary (e.g., \ from general to specialized), and
the information represented in a collection may cover a broad range of topics.
Lexical approaches, e.g., \ TF.IDF or BM25 \cite{bm25}, have dominated textual IR for many years. Mainly due to manageable computational cost, but still offering 
decent performance. 
Recently, a strong trend has been observed towards developing neural retriever models that should outperform lexical approaches.
Pretrained language models like BERT \cite{devlin-etal-2019-bert} appeared to be a good basis for dense retrieval approaches. Bi-encoder architecture as presented in dense passage retriever (DPR) \cite{DBLP:journals/corr/abs-2004-04906} and sentence BERT \cite{reimers2019sentence} are commonly used and express high performance, especially on in-domain datasets. The query and document are represented by single vectors, which facilitates applications of fast vector databases i.e.\ FAISS \cite{faiss}. The main drawback of such models is their lower performance on out-of-domain data. On the other hand, the BM25 approach achieves better results in such scenario. A potential approach involves utilising a multi-vector representation of the query and document, as exemplified in ColBERT \cite{colbert}. These approaches utilise the late interaction paradigm. ColBERT encodes documents and queries in multiple vectors, where each output vector corresponds to the input token. During inference time, ColBERT computes the Cartesian product between queries and documents, which can enhance retrieval outcomes, but also necessitates storing a huge index in memory. 
To improve performance of single vector representations, it was shown that language models have structural readiness, and it is possible to pre-train the model towards bi-encoder structure \cite{gao-callan-2021-condenser}. Also, we can explicitly aggregate sentence-level representation
with token representation, which is obtained using a weight matrix from Mask Language Modeling (MLM) pre-training task and treating all input sentence tokens as a probability distribution over model dictionary \cite{aggretriever}. The aggregated tokens representation, in conjunction with CLS vector, is fine-tuned to the retrieval task.

\subsection{Unsupervised pretraining}

Unsupervised methods are mainly aimed at zero-shot schemes. 
In IR, most methods focus on data augmentation and generation of pseudo queries.
Inverse Cloze Task (ICT) \cite{DBLP:journals/corr/abs-1906-00300} 
resembles finding evidence to answer a question. In contrast to the standard Cloze task -- predicting masked text given its context -- the ICT requires anticipating the context given a sentence. The unsupervised equivalent of the question-evidence pair is the sentence-context pair -- the context of a sentence has semantic meaning and can be used to infer information that is not directly expressed in the sentence. 
Another method of building positive query-document pair instances  without supervision is dropout as a Positive Instance (DaPI) from  SimCSE \cite{sim_ce_dapi}. 
To perturb the input sentence representation, dropout is applied to transformer models' fully-connected layers and attention probabilities. As a result, an obtained representation can be treated as a positive instance pair with the same sentence but different hidden dropout mask.
The promising performance of both methods and evaluation on English BEIR benchmark was shown in LaPraDor \cite{xu-etal-2022-laprador}.

\subsection{Passage Re-ranking}

BERT \cite{devlin-etal-2019-bert} enabled approaches based on cross-encoders, e.g., \ \cite{reimers2019sentence}, in which we obtain a joint embedding of a document and an input query, on the token level. In this approach, BERT processes a document and a query simultaneously, scoring their relationship. 
Due to computational cost, cross-encoders are particularly popular in two-stage retrieval architectures. The first stage extracts the most relevant documents with a light and fast model (e.g., \ BM25 \cite{bm25}). Cross-encoders are used in the next stage for re-ranking. A reranker, e.g., \ a cross-encoder, recomputes document scores from the first stage (see Figure \ref{fig:retriev-rerank}). Alternatively, generative sequence-to-sequence language models were also proposed for re-ranking. MonoT5 \cite{nogueira-etal-2020-document} is an adaptation of the T5 model \cite{raffel2020exploring} for IR task. The model had achieved state-of-the-art results in zero-shot setting. The sequence-to-sequence model is triggered by a prompt containing a query followed by a document. The model is expected to assess their relevance by producing ``true'' or ``false'' token in a generative way. The idea of two-stage approach is presented in Figure~\ref{fig:retriev-rerank}. Initially, a set of top K documents is retrieved using techniques such as BM25. Subsequently, the retrieved documents are reranked based on the query using a reranker.


\subsection{BEIR benchmark}

BEIR is a benchmark for zero-shot IR encompassing various tasks -- their sizes are shown in Table~\ref{table:beir_stats}. The authors of BEIR benchmark 
aimed at obtaining a large-scale data collection representing diversified IR tasks, 
with various features of text data and queries, e.g. collecting queries and documents of different lengths and style, also originating from different domains, not only news or Wikipedia articles. 
Different domains are meant to represent real-world data settings and should be challenging for various methods. Moreover, the datasets were annotated by utilising different annotation strategies, e.g. performed by crowd-workers but also experts in specific cases.

\begin{table}
    \scriptsize

    \begin{center}
    \begin{tabular}{wl{1.2cm}wr{1.2cm}wr{1.2cm}rrr} %
         \textbf{Dataset} & \textbf{\#Test queries} & \textbf{Corpus size}  & \textbf{Avg. Q Len} & \textbf{Avg. D Len}\\
         MSMARCO        & 6980 &   8.8M & 5.33 & 49.63 & \\[1pt]
         TREC-COVID     &    50 &   171K & 9.44 & 137.05 & \\[1pt]
         NFCorpus       &   323 &   3.6K & 3.37 & 205.96 & \\[1pt]
         NQ             & 3 452 &  2.68M & 7.33 & 66.89 & \\[1pt]
         HotpotQA       & 7 405 &   5.2M & 15.64 & 38.67 & \\[1pt]
         FiQA           &   648 &    57K & 9.76 & 113.96 & \\[1pt]
         ArguAna        & 1 406 &     9K & 168.01 & 142.48 & \\[1pt]
         Touche-2020    &    49 &   382K & 7.12 & 125.48 & \\[1pt]
         CQADupstack    & 13 145&   547K & 7.86 & 110.76 & \\[1pt]
         Quora          & 10 000&   523K & 8.13 & 9.85 & \\[1pt]
         DBPedia        &   400 &  4.63M & 4.82 & 41.61 & \\[1pt]
         SciDocs        & 1 000 &    25K & 9.70 & 150.15 & \\[1pt]
         SciFact        &   300 &     5K & 11.74 & 187.66 & \\[1pt]
    \end{tabular}
    \end{center}
    \caption{\label{dataset--number-description} Number of queries, corpus size, average (mean) query and document word length across all datasets in BEIR-PL benchmark.}
    \label{table:beir_stats}
\end{table}

\section{Methodology}

In this section, we present the steps taken to create BEIR-PL benchmark dataset. 
As our aim was to build a large-scale benchmark as a reference point for comparing different IR models in Polish, we decided to translate the entire BEIR benchmark using automated Machine Translation. Subsequently, we trained and evaluated baseline models on the newly created resources. Baseline models will be publicly available to the research community. The selection of baseline models was dictated by recent advances in dense information retrieval and reranking models existing in the literature. 

\subsection{Translation of the datasets}

\begin{table}[htbp]
\scriptsize
\begin{center}
\setlength{\tabcolsep}{3pt}
\setlength\belowcaptionskip{4pt}
\begin{tabular}{llcccccccccc}
   \toprule
   \small{\textbf{{\scriptsize }}} & & 
   \rotatebox{90}{\scriptsize{DBPedia Q.}} &
   \rotatebox{90}{\scriptsize{FiQa Q.}} & \rotatebox{90}
   {\scriptsize{HotpotQA Q.}} &
   \rotatebox{90}
   {\scriptsize{HotpotQA C.}} &
   \rotatebox{90}{\scriptsize{MSMARCO C.}} & \rotatebox{90}{\scriptsize{MSMARCO Q.}} &
   \rotatebox{90}{\scriptsize{SciFact Q.}}
   &
   \rotatebox{90}{\scriptsize{Quora Q.}}
   &
   \rotatebox{90}{\scriptsize{Quora C.}}\\ 
   \midrule
   {\scriptsize LaBSE} &  &  0.89 & 0.88 & 0.90 & 0.93 & 0.91 & 0.87 & 0.88 & 0.90 & 0.91  \\[2pt]
   \midrule
   {\scriptsize Semantic} & & 0.85 & 0.80 & 0.84 & 0.88 & 0.84 & 0.91 & 0.80 & 0.82 & 0.92  \\[2pt]
   {\scriptsize Strict}   & & 0.61 & 0.52 & 0.72 & 0.70 & 0.52 & 0.77 & 0.72 & 0.72 & 0.78 \\[3pt]

   \bottomrule
\end{tabular}
\caption{We have randomly chosen samples from datasets and assessed their translation correctness accuracy in two settings. \textit{Semantic} setting is when the translated query or document is understandable and adequate to the IR task, in a way that it does not change the meaning of the original text. \textit{Strict} setting is the translation assessment done by a professional linguist in context of IR task evaluating both the semantic consistency, style and relevance of terminology. We also provide automated text similarity score measured by multilingual language model -- LaBSE. We evaluated: queries - Q. and passages from corpus - C. }
\label{tab:translation_eval}
\end{center}
\end{table}

To create a large-scale resource for information retrieval, it is necessary to obtain a significant number of annotated query-passage pairs. However, the high cost of the annotation procedure can make this infeasible. Additionally, linguistic translation from foreign languages over millions of documents is both demanding and costly. As a result, machine translation can serve as a cost-effective solution to enrich resources in low-resource languages such as Polish.
In order to process and translate the available BEIR benchmark's datasets into the Polish language, we used Google Translate service. This service has been previously used to translate mMarco \cite{DBLP:journals/corr/abs-2108-13897} dataset into various languages, but unfortunately, the Polish language was not included in this study. It has been shown during the development of mMarco,
that the results obtained from Google Translate were better than the translation by the available open-source machine translation model, namely Helsinki \cite{tiedemann-thottingal-2020-opus}, which can be downloaded from the HuggingFace\footnote{\url{http://huggingface.co}} repository. For that reason, we decided to rely on Google Translate service.

Queries are pre-defined natural language questions used to evaluate the performance of an IR system, while corpus refers to the set of documents that the system searches to find answers to the queries. Qrels, on the other hand, represent the relevance judgments indicating the relationships between the queries and documents in the corpus. Queries and corpus are defined and stored in a JSONL format 
and qrels in tsv format. That ensures that our resource can be treated as an extension to and be fully compatible with the multilingual BEIR benchmark in the future.

The size of the obtained resource, as illustrated in Table~\ref{dataset--number-description} causes that manual verification of it would be very laborious. Instead, in order to search for potential problems, we have selected 100 random queries and passages which were evaluated by a linguist in a Strict setting and a researcher in Semantic setting as shown in Table \ref{tab:translation_eval}. Moreover, multilingual contextual embeddings model, namely LaBSe \cite{feng-etal-2022-language}, has been used to compare source texts and their translations in automatic manner.
We could observe high semantic similarity reported by LaBSe model and selective manual inspection proved that that most of the translations were adequate to the IR task, but not perfect. Errors were particularly noticed in the translation of Named Entities and when translated, queries sought the same information but had incorrect phrasing.

\begin{figure*}[htbp]
    \centering
    \subfloat[][BM25 performance evaluated using \\ recall@1K measure.]{\resizebox{0.5\textwidth}{!}{
    \begin{tikzpicture}
    \begin{axis}[
        ybar,
        ymax=1.1,
        ymin=0,
        width  = 10cm,
        height = 4.25cm,
        bar width=8pt,
        ylabel={Recall@1K},
        nodes near coords,
        xticklabel style={rotate=80},
        ylabel style={font=\scriptsize},
        xtick={1,2,3,4,5,6,7,8,9,10},
        xmin=0,
        xmax=11,
        table/header=false,
        table/row sep=\\,
        every node near coord/.append style={rotate=90, anchor=west, font=\scriptsize},
        xticklabels from table={\footnotesize
          English\\\footnotesize Spanish\\\footnotesize French\\\footnotesize Italian\\
          \footnotesize
          Portuguese\\\footnotesize Indonesian\\\footnotesize German\\\footnotesize Russian\\
          \footnotesize
          Chinese\\\footnotesize Polish\\
          }{[index]0},
        enlarge y limits={value=0.2,upper}
    ]
    \addplot[ybar,fill=gray!30!black,draw=black] plot coordinates{
    (1,0.857)
    (2,0.770)
    (3,0.769)
    (4,0.753)
    (5,0.744)
    (6,0.767)
    (7,0.674)
    (8,0.685)
    (9,0.678)};
    \addplot[ybar,fill=red!50!black,draw=black] plot coordinates{(9.5,0.667)};
    \end{axis}
    \coordinate(bbs)at(current bounding box.south);
\end{tikzpicture}
    }}
    \subfloat[][BM25 performance evaluated using \\ MRR@10 measure.]{\resizebox{0.5\textwidth}{!}{
         \begin{tikzpicture}
     \begin{axis}[
        ybar,
        ymin=0,
        ymax=0.3,
        width  = 10cm,
        height = 4.28cm,
        bar width=8pt,
        ylabel={MRR@1K},
        nodes near coords,
        xticklabel style={rotate=80},
        ylabel style={font=\scriptsize},
        tick label style={/pgf/number format/fixed},
        xtick={1,2,3,4,5,6,7,8,9,10},
        xmin=0,
        xmax=11,
        table/header=false,
        table/row sep=\\,
        every node near coord/.append style={rotate=90, anchor=west, font=\scriptsize},
        xticklabels from table={\footnotesize
          English\\\footnotesize Spanish\\\footnotesize French\\\footnotesize Italian\\
          \footnotesize
          Portuguese\\\footnotesize Indonesian\\\footnotesize German\\\footnotesize Russian\\
          \footnotesize
          Chinese\\\footnotesize Polish\\
          }{[index]0},
        enlarge y limits={value=0.2,upper}
    ]
    \addplot[ybar,fill=gray!30!black,draw=black] plot coordinates{
    (1,0.184)
    (2,0.158)
    (3,0.155)
    (4,0.153)
    (5,0.152)
    (6,0.149)
    (7,0.136)
    (8,0.124)
    (9,0.116)};
    \addplot[ybar,fill=red!50!black,draw=black] plot coordinates{(9.5,0.120)};;
    \end{axis}
    \coordinate(bbs)at(current bounding box.south);
\end{tikzpicture}
    }}
    \caption{BM25 performance on MS~Marco passage retrieval on different languages\cite{DBLP:journals/corr/abs-2108-13897}.}
    \label{fig:bm25_multilingual_results}
\end{figure*}
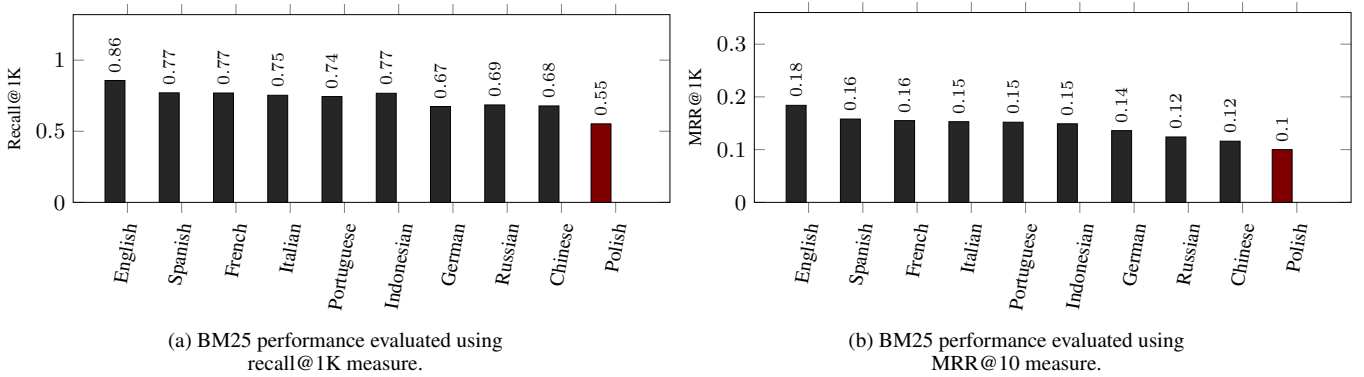

\subsection{Baseline models}

This section briefly describes the baseline IR models we used for our evaluation study. The main baseline was computed using lexical matching with the BM25 implementation from Elasticsearch engine\footnote{\url{https://www.elastic.co/}} with Stempel Polish analysis plugin. This is a standard baseline method used in IR, which has demonstrated strong performance and computational efficiency across various domains. It is also typically used in the first stage of retrieval, as shown in Figure \ref{fig:retriev-rerank}. The baseline neural models can be divided into following categories:

\begin{itemize}
    \item Dense retrievers -- these models are repesenting passages with single vectors and can substitute BM25 matching. We evaluated three BERT-only bi-encoder models: the unsupervised bi-encoder based on ICT technique \cite{DBLP:journals/corr/abs-1906-00300} with HerBERT \cite{mroczkowski-etal-2021-herbert} as its core and pre-existing multilingual model LaBSE \cite{feng-etal-2022-language}.
    \item Rerankers -- these models can be used only as rerankers due to their computational inefficiency. We evaluated two HerBERT-based models and two T5-based models \cite{raffel2020exploring}. The models were pre-trained for Polish language on translated MSMARCO data. Finally, we evaluated one late-interaction reranker based on ColBERT architecture \cite{colbert}. Also, included already existing pretrained multilingual model mMiniLM \cite{DBLP:journals/corr/abs-2108-13897} reranker.
\end{itemize}



\begin{table*}[tbp]
\begin{center}
\setlength{\tabcolsep}{3pt}
\setlength\belowcaptionskip{-20pt}
\begin{tabular}{lcccccccccccc}
   \small{\textbf{{\scriptsize Metric}}} & & \rotatebox{90}{\scriptsize{MSMARCO}} &
   \rotatebox{90}{\scriptsize{TREC-COVID}} & \rotatebox{90}{\scriptsize{NFCorpus}} &
   \rotatebox{90}{\scriptsize{\texttt{NQ}}} & \rotatebox{90}{\scriptsize{HotpotQA}} &
   \rotatebox{90}{\scriptsize{FiQA}} & \rotatebox{90}{\scriptsize{ArguAna}} 
   & \rotatebox{90}{\scriptsize{CQADupstack}} 
   & \rotatebox{90}{\scriptsize{DBPedia}} &
   \rotatebox{90}{\scriptsize{SciDocs}} & \rotatebox{90}{\scriptsize{SciFact}} \\
   NDCG@10 & \texttt{PL}   & 41.9 & 61.0 & 31.9 & 20.1 & 49.2 & 19.0 & 41.36 & 28.37 & 22.9 & 14.1 & 62.5 \\[3pt]
   NDCG@10 & \texttt{EN}   & 47.7 & 68.9 & 34.3 & 32.6 & 60.2 & 25.4 & 47.2 & 32.5 & 32.1 & 16.5 & 69.1 \\[3pt]
   Recall @100 & \texttt{PL} & 34.6 & 10.1 & 24.6 & 57.9 & 67.1 & 44.1 & 93.5 & 53.9 & 30.1 & 33.0 & 88.4 \\[3pt]
   Recall @100 & \texttt{EN} & 45.0 & 11.7 & 26.0 & 78.3 & 76.3 & 54.9 & 95.2 & 62.1 & 43.5 & 36.8 & 92.0 \\
\end{tabular}
\end{center}
\caption{An overall comparison between results achieved on BEIR-PL and original BEIR using lexical BM25 matching on test data. The results are evaluated using NDCG @10 and Recall @100 metrics.}
\label{tabel:bm25_multilingual}
\end{table*}

\subsubsection{Unsupervised dense bi-encoder}
To evaluate the unsupervised models and check how well they are performing on the benchmark data, we decided to fine-tune the HerBERT-base model \cite{mroczkowski-etal-2021-herbert} with ICT unsupervised tasks on BEIR-PL benchmark datasets. For each document, a pseudo-query was generated and used as a training positive instance. We utilized the model as a bi-encoder, in which it encodes queries and documents independently into single dense vector representations. Those vector representations can be compared using cosine similarity and saved to create a dense index of documents. 

\subsubsection{HerBERT based reranker models}
We further evaluated reranker models in a setting where the top 100 search results are retrieved by BM25 are presented as an input to the model. The model output is the re-ranked order of documents corresponding to the query. We trained HerBERT-base and HerBERT-large rerankers on the BEIR-PL MS~MARCO dataset.

\subsubsection{Polish T5 based reranker models}
Furthermore, we trained and evaluated sequence-to-sequence MonoT5 rerankers based on plT5 language models \cite{chrabrowa2022evaluation}, in both base and large variants. We used special tokens \emph{\_prawda} (`true') and \emph{\_fałsz} (`false'), to represent positive and negative relevance between query and passage. This architecture is composed of an encoder and decoder, which may lead to different performance compared to HerBERT, which is a BERT-based model \cite{devlin-etal-2019-bert}. 

\subsection{Late interaction HerBERT based reranker model}
Finally, we trained and evaluated late-interaction model named ColBERT \cite{colbert}, with HerBERT base as its core language model. The maximum document length was set to 180 tokens in our configuration. Creating an index for the retrieval task using ColBERT requires large disk and memory space -- the model stores in memory the tokens obtained from corpus encoding (the index size of MSMARCO-PL is estimated to be at least 200GB). Due to that reason, we decided to use ColBERT as a reranker, which does not require creating enormous indexes.

\subsection{Pre-existing multilingual models}
We compared our models with already available multilingual models, to check their performance on Polish language.

As a bi-encoder, we tested an already pre-trained multilingual LaBSE model. This model was fine-tuned to the sentence embedding task and showed competitive results compared with other fine-tuned multilingual retrievers described on the SentenceTransformers page\footnote{\url{https://www.sbert.net/}}.

In case of fine-tuned multilingual reranker model, we decided to test mMiniLM model \cite{DBLP:journals/corr/abs-2108-13897} which was shared by cross-encoder organization on Hugging Face platform. The model was fine-tuned to re-rank queries and documents from the multilingual MS~Marco corpora and was trained on 80M training examples.


\subsection{Experimental setup}
Our models (all except ColBERT) were trained on two NVIDIA--RTX3090 GPUs with 24GB of memory each. The unsupervised HerBERT-base ICT bi-encoder was trained for 203 hours with batch size 64 for about 1.8M iterations. The learning rate hyperparameter was set to $2e^{-4}$. 
For fine-tuned variants of HerBERT-based rerankers, we trained the models for $\approx$25 hours with a batch size of 32. The HerBERT-base model was trained on 20M examples, and the HerBERT-large model on 3.2M. Both T5-based models were trained for ~20 hours with gradient accumulation steps set to 16 and batch size 16. The T5-base model was trained on 5M examples, and the T5-large model on 645K examples. The difference in training size between the base and large models was caused by the computational cost of model training. 
 ColBERT model was trained on a single NVIDIA GeForce RTX 2080 Ti GPU with 12GB of memory. The model was trained for 102h with batch size 8 and learning rate set to $3e^{-6}$. Moreover, the max document token number was set to 180 for documents and 32 for queries.



\subsection{Evaluation metrics}

For the comparison of the models, we applied the most commonly used metrics in IR research: Mean Reciprocal Rank(MRR), Normalised Cumulative Discount Gain (nDCG), and Recall. More details about the metrics in the Appendix \ref{appedix_metric}.

\begin{table*}[htbp]
\scriptsize
\begin{center}
\setlength{\tabcolsep}{2.8pt}
\setlength\belowcaptionskip{-20pt}
\begin{tabular}{lllccccccccccccc}
   \toprule
   \small{\textbf{{\scriptsize Model}}} & & &
   \rotatebox{90}{\scriptsize{MSMARCO}} &
   \rotatebox{90}{\scriptsize{TREC-COVID}} & \rotatebox{90}{\scriptsize{NFCorpus}} &
   \rotatebox{90}{\scriptsize{NQ}} & \rotatebox{90}{\scriptsize{HotpotQA}} &
   \rotatebox{90}{\scriptsize{FiQA}} & \rotatebox{90}{\scriptsize{ArguAna}} & \rotatebox{90}{\scriptsize{Touche-2020}} 
   & \rotatebox{90}{\scriptsize{CQADupstack}} & \rotatebox{90}{\scriptsize{Quora}} 
   & \rotatebox{90}{\scriptsize{DBPedia}} &
   \rotatebox{90}{\scriptsize{SciDocs}} & \rotatebox{90}{\scriptsize{SciFact}} \\  
   \midrule
   {\scriptsize BM25} & \multirow{9}{*}{\rotatebox{90}{{\scriptsize NDCG@10}}} & \xmark &  41.92 & 61.00 & 31.92 & 20.14 & 49.21 & 19.00 & \textbf{41.36} & \textbf{32.89} & 28.37 & 63.87 & 22.98 & 14.13 & 62.56 \\[3pt]
   {\scriptsize ICT}  & & \cmark & 29.02 & 22.48 & 15.05 & 0.79 & 7.97 & 9.61 & 20.82 & 13.07 & 9.98 & 72.61 & 14.74 & 3.60 & 38.17 \\[3pt]
   {\scriptsize LaBSE}& & \xmark & 17.55 & 18.52 & 17.45 & 9.62 & 19.74 & 7.66 & 38.56 & 5.16 & 20.01 & 75.00 & 16.08 & 7.47 & 39.79 \\[3pt]
   \cline{3-16}\\[-1ex]
   {\scriptsize mMiniLM}           & & \xmark & 64.10 & 64.70 & 29.84 & 34.16 & 59.81 & 27.67 & 22.10 & 30.36 & 27.94 & 68.73 & 29.61 & 12.21 & 60.08 \\[3pt]
   {\scriptsize HerBERT}$_{base}$  & & \cmark & 62.45 & 56.63 & 30.17 & 33.53 & 57.44 & 26.33 & 21.09 & 26.76 & 26.85 & 68.74 & 27.67 & 12.01 & 60.31  \\[3pt]
   {\scriptsize HerBERT}$_{large}$ & & \cmark & 64.00 & 58.44 & 27.77 & 36.94 & 61.78 & 28.95 & 33.76 & 27.79 & 28.53 & 73.90 & \textbf{30.00} & 13.03 & 54.13 \\[3pt]
   \cline{3-16}\\[-1ex]
   {\scriptsize T5}$_{base}$      & & \cmark & \textbf{64.23} & 68.53 & 32.79 & 35.77 & \textbf{63.13} & 25.93 & 15.52 & 23.97 & 30.40 & 49.94 & 18.21 & 14.35 & 68.54 \\[3pt]
   {\scriptsize T5}$_{large}$      & & \cmark & 62.81 & \textbf{73.66} & \textbf{35.61} & \textbf{39.53} & 62.21 & \textbf{30.92} & 13.04 & 26.21 & \textbf{32.05} & 50.82 & 18.20 & \textbf{14.79} & \textbf{69.55} \\[3pt]
   \cline{3-16}\\[-1ex]
   {\scriptsize ColBERT}       & & \cmark & 60.11 & 64.72 & 31.79 & 31.84 & 51.85 & 25.82 & 37.10 & 26.76 & 27.92 & \textbf{75.44} & 26.87 & 11.51 & 55.85 \\[3pt]
   \midrule
   {\scriptsize BM25} & \multirow{9}{*}{\rotatebox{90}{{\scriptsize MRR@10}}} & \xmark &  68.09 & 84.03  & 46.44 & 17.50 & 64.18 & 23.73 & \textbf{33.17} & \textbf{68.18} & 27.90 & 63.75 & 48.56 & 25.59 & 62.56 \\[3pt]
   {\scriptsize ICT} & & \cmark & 43.91 & 39.39 & 29.81 & 0.72 & 10.46 & 11.63 & 13.53 & 28.88 & 9.52 & 71.77 & 31.05 & 10.23 & 6.12 \\[3pt]
   {\scriptsize LaBSE} & & \xmark & 40.90 & 39.23 & 31.99 & 8.27 & 27.12 &  9.36 & 31.85 & 13.65 & 19.49 & 74.39 & 36.53 & 13.37 & 36.45 \\[3pt]
   \cline{3-16}\\[-1ex]
   {\scriptsize mMiniLM}           & & \xmark & 91.55 & 85.81 & 44.37 & 31.47 & 77.39 & 35.57 & 17.43 & 60.19 & 27.95 & 66.80 & 56.96 & 22.39 & 56.79 \\[3pt]
   {\scriptsize HerBERT}$_{base}$ & & \cmark & 90.11 & 79.66 & 46.78 & 30.74 & 73.98 & 34.28 & 16.31 & 50.54 & 26.42 & 67.04 & 52.02 & 21.34 & 57.37  \\[3pt]
   {\scriptsize HerBERT}$_{large}$ & & \cmark & \textbf{91.86} & 78.80 & 42.50 & 34.40 & 79.41 & 37.29 & 18.31 & 45.23 & 27.90 & 73.06 & \textbf{58.42} & 23.21 & 58.63 \\[3pt]
   \cline{3-16}\\[-1ex]
   {\scriptsize T5}$_{base}$     & & \cmark & 89.53 & 80.40 & 48.62 & \textbf{38.63} & \textbf{80.60} & 33.77 & 13.11 & 43.20 &  30.03 & 49.32 & 38.22 & 26.50 & 66.62 \\[3pt]
   {\scriptsize T5}$_{large}$    & & \cmark & 89.15 & \textbf{90.73} & \textbf{51.23} & 36.67 & 79.59 & \textbf{39.75} & 11.09 & 46.15 & \textbf{31.70} &  50.19 & 39.96 & \textbf{26.92} & \textbf{67.73} \\[3pt]
   \cline{3-16}\\[-1ex]
   {\scriptsize ColBERT}      & & \cmark & 85.89 & 81.07 & 47.24 & 29.00 & 68.67 & 32.93 & 30.06 & 49.26 & 27.76 & \textbf{75.34} & 51.64 & 20.79 & 52.78 \\[3pt]
   \bottomrule
\end{tabular}
\caption{A complete comparison of Polish retrieval and reranking models in Information Retrieval task. We present evaluation results calculated for all data subsets of BEIR-pl benchmark. We evaluated a wide range of IR models and we compared their performance with BM25 lexical matcher. To evaluate the performance we used NDCG @10 and MRR @10 evaluation metrics. The ICT model is an unsupervised bi-encoder HerBERT-base dense network. \cmark denotes Polish models that were fine-tuned in this study, where \xmark denotes already fine-tuned multilingual models we evaluated}
\label{tab:general}
\end{center}
\end{table*}

\begin{table*}[htbp]
\scriptsize
\begin{center}
\setlength{\tabcolsep}{3pt}
\setlength\belowcaptionskip{-20pt}
\begin{tabular}{lllcccccccccccc}
   \toprule
   \small{\textbf{{\scriptsize Model}}} & & &
   \rotatebox{90}{\scriptsize{android}} &
   \rotatebox{90}{\scriptsize{english}} & \rotatebox{90}{\scriptsize{gaming}} &
   \rotatebox{90}{\scriptsize{gis}} & \rotatebox{90}{\scriptsize{math}} &
   \rotatebox{90}{\scriptsize{physics}} & \rotatebox{90}{\scriptsize{program.}} & \rotatebox{90}{\scriptsize{stats}} 
   & \rotatebox{90}{\scriptsize{tex}}
   & \rotatebox{90}{\scriptsize{unix}} &
   \rotatebox{90}{\scriptsize{webmasters}} & \rotatebox{90}{\scriptsize{wordpress}} \\  
   \midrule
   {\scriptsize BM25} & \multirow{9}{*}{\rotatebox{90}{{\scriptsize NDCG@10}}} & \xmark &    36.29 & 25.34 & 38.58 & 25.95 & 19.06 & 30.97 & 32.21 & 28.42 & 20.94 & 26.86 & 30.61 & 25.29 \\[2pt]
   {\scriptsize ICT}              & & \cmark & 14.45 & 11.21 & 19.96 & 5.70 & 5.90 & 11.49 & 12.02 & 6.46 & 4.91 & 9.54 & 12.58 & 5.62\\[3pt]
   {\scriptsize LaBSE}            & & \xmark & 30.31 & 15.55 & 31.50 & 17.59 & 13.62 & 22.50 & 19.98 & 17.11 & 13.20 & 21.01 & 22.60 & 15.05 \\[3pt]
   \cline{3-15}\\[-1ex]
   {\scriptsize mMiniLM}           & & \xmark & 33.92 & 28.11 & 39.40 & 25.73 & 17.85 & 30.64 & 28.72 & 24.88 & 22.43 & 27.38 & 29.79 & 26.45 \\[3pt]
   {\scriptsize HerBERT}$_{base}$ & & \cmark & 33.99 & 26.80 & 38.20 & 23.09 & 17.29 & 30.92 & 28.22 & 23.42 & 21.24 & 26.25 & 28.13 & 24.35 \\[3pt]
   {\scriptsize HerBERT}$_{large}$ & & \cmark & 37.72 & \textbf{30.24} & 38.20 & 23.09 & 18.48 & 30.92 & 29.63 & 24.57 & 22.37 & 30.58 & 31.03 & 25.64 \\[3pt]
   \cline{3-15}\\[-1ex]
   {\scriptsize T5}$_{base}$      & & \cmark & 39.06 & 14.17 & 45.86 & 29.14 & 21.50 & 36.14 & 32.90 & 29.81 & 23.83 & 30.77 & 33.65 & 28.04 \\[3pt]
   {\scriptsize T5}$_{large}$      & & \cmark & \textbf{42.05} & 14.54 & \textbf{48.30} & \textbf{30.75} & \textbf{23.02} & \textbf{37.68} & \textbf{34.68} & \textbf{30.90} & \textbf{24.78} & \textbf{32.08} & \textbf{35.66} & \textbf{30.18} \\[3pt]
   \cline{3-15}\\[-1ex]
   {\scriptsize ColBERT}           & & \cmark & 35.62 & 27.84 & 39.61 & 25.27 & 16.92 & 31.03 & 29.54 & 24.05 & 21.65 & 27.47 & 30.33 & 25.74 \\[3pt]
   \midrule
   {\scriptsize BM25} & \multirow{9}{*}{\rotatebox{90}{{\scriptsize MRR@10}}} & \xmark & 35.90 & 26.39 & 37.86 & 23.95 & 18.76 & 31.42 & 32.66 & 27.59 & 20.88 & 26.28 & 29.29 & 23.82 \\[3pt]
   {\scriptsize ICT}              & & \cmark & 14.51 & 11.32 & 18.65 & 4.82 & 5.40 & 10.88 & 11.40 & 6.12 & 4.78 & 9.03 & 12.57 & 4.80 \\[3pt]
   {\scriptsize LaBSE}            & & \xmark & 30.12 & 16.07 & 30.29 & 15.85 & 13.14 & 22.71 & 19.84 & 16.29 & 13.15 & 20.73 & 22.11 & 13.68 \\[3pt]
   \cline{3-15}\\[-1ex]
   {\scriptsize mMiniLM}           & & \xmark & 34.86 & 29.66 & 39.01 & 24.60 & 17.87 & 31.15 & 29.38 & 23.98 & 22.93 & 27.24 & 29.05 & 25.67 \\[3pt]
   {\scriptsize HerBERT}$_{base}$ & & \cmark & 33.50 & 27.82 & 37.04 & 21.52 & 17.31 & 30.82 & 28.69 & 22.49 & 21.39 & 26.08 & 26.97 & 23.46 \\[3pt]
   {\scriptsize HerBERT}$_{large}$ & & \cmark & 38.23 & \textbf{31.22} & 37.04 & 21.52 & 18.49 & 30.82 & 30.41 & 23.27 & 22.47 & 26.06 & 30.52 & 24.79 \\[3pt]
   \cline{3-15}\\[-1ex]
   {\scriptsize T5}$_{base}$      & & \cmark & 39.23 & 14.48 & 45.05 & 27.27 & 21.22 & 36.70 & 33.05 & 28.94 & 23.75 & 30.72 & 33.22 & 26.82 \\[3pt]
   {\scriptsize T5}$_{large}$     & & \cmark & \textbf{42.35} & 14.75 & \textbf{47.69} & \textbf{28.79} & \textbf{22.86} & \textbf{38.40} & \textbf{34.78} & \textbf{29.94} & \textbf{24.69} & \textbf{31.86} & \textbf{35.61} & \textbf{28.79} \\[3pt]
   \cline{3-15}\\[-1ex]
   {\scriptsize ColBERT}           & & \cmark & 35.79 & 28.90 & 38.96 & 24.17 & 17.25 & 31.44 & 29.66 & 23.35 & 22.09 & 27.30 & 29.84 & 24.40 \\[3pt]
   \bottomrule
\end{tabular}
\caption{A comparison of retriever and reranker models evaluated on data subsets from CQDupstack dataset. To evaluate the models we used NDCG @10 and MRR @10 evaluation metrics.}
\label{tab:cqadupstack}
\end{center}
\end{table*}

\section{Results and Discussion}

BM25 -- lexical level retrieval algorithm --  performs better for the English language than for Polish, as showed in Table~\ref{tabel:bm25_multilingual}. Moreover, it demonstrates one of the lowest scores for Polish when compared with the multilingual MS~Marco dataset results among all different languages, as shown in Figure~\ref{fig:bm25_multilingual_results}.
The main cause of such a low performance scores for Polish language is that Polish is a highly inflected language with large number of word forms per lexeme (also Proper Names are inflected) and a complex morphological structure. In such a case, lexical matching is less effective than in the case of other languages. 

On the other hand, our results show that the BM25 baseline is a strong baseline for neural models, even in the case of Polish, as shown in Table~\ref{tab:general} and Table~\ref{tab:cqadupstack}.
When compared with bi-encoder models, i.e., \ the unsupervised ICT model initialised with HerBERT-base and the LaBSE model pre-trained for sentence semantic similarity, BM25 is a better choice for most datasets as is presented in Figure~\ref{fig:retriev-rerank}.

Interesting results have been obtained for the Quora dataset, in which bi-encoders have achieved very high results. The underlying reason for this observation is that the Quora dataset primarily concentrates on the task of determining whether a question has been duplicated. This requires from the models to rely on semantic similarity of sentences, that  was indirectly a part of pre-training procedure for the LaBSE model. The ICT fine-tuning seems to enhance the performance of the model for this particular task effectively, which could be attributed to the model's exposure to similar questions from the corpus with minor modifications during the training procedure. The ColBERT model was able to enhance the rankings, outperforming other models. This suggests that the ColBERT model is not only effective in retrieving passages with answers to queries but also excels in general semantic similarity.

The performance of the ICT bi-encoder model is surprisingly low on Natural Questions (NQ) and HotpotQA datasets, which may be due to the complexity of these datasets. Both datasets are derived from the Question Answering task. Questions might be very distant regarding lexical similarity from the retrieved documents, and the ICT task is an insufficient approach in this case. For those datasets, a deeper understanding of provided text is essential.

The results of the reranker models show a significant improvement over BM25 lexical matching, which means neural cross-encoders can re-arrange retrieval results into a better pertinence ranking. Only in the case of the ArguAna and Touche-2020 datasets, the performance of reranker models, is lower than BM25 results. The objective of the IR task in the ArguAna dataset is to find a counterargument for a provided argument. This might not be straightforward for reranker models, given the difference in the meanings of the sentences being compared. Consequently, a lower performance in this task might actually indicate a superior model in terms of text comprehension. Similarly, the Touche-2020 dataset focuses on locating arguments related to a specific topic and might also include counterarguments to the thesis in the query.

The sequence-to-sequence T5 models express a slight improvement over BERT-based rerankers on most datasets, but there are cases where we can notice a significant drop in performance. We believe that, in the case of ArguAna dataset, the reason for the diminished performance can be attributed to the heightened sensitivity of sequence-to-sequence models to differences in semantic meaning between arguments and counterarguments, as previously mentioned. Furthermore, the results on Quora dataset are worse after re-ranking than BM25, indicating that the task is not inherently intuitive for this T5 model class after fine-tuning on MS MARCO, particularly in contrast to HerBERT-based rerankers. Also, T5-large model achieves the best performance on all CQDupstack subsets, except English, as shown in Table~\ref{tab:cqadupstack}.

The late-interaction ColBERT model demonstrates significant improvements over the BM25 retriever on the majority of datasets. Considering that it is solely a late-interaction model and not a full cross-encoder, one might anticipate its performance to be inferior to that of the HerBERT reranker. However, there are instances where the actual performance surpasses expectations, with slightly higher results on certain datasets, namely Quora, TREC-COVID and NFCorpus. This observation could be attributed to the fact that both TREC-COVID and NFCorpus are medical-related datasets, wherein the retrieval task might be simplified to identifying keywords present in both the query and the passage.

\begin{table}[htbp]
\scriptsize
\begin{center}
\setlength{\tabcolsep}{3pt}
\setlength\belowcaptionskip{-20pt}
\begin{tabular}{llcccccc}
   \toprule
   \small{\textbf{{\scriptsize Model}}} & & 
   \rotatebox{90}{\scriptsize{Allegro-Faq Test A}} &
   \rotatebox{90}{\scriptsize{Allegro-Faq Test B}} & \rotatebox{90}{\scriptsize{Legal Questions Test A}} &
   \rotatebox{90}{\scriptsize{Legal Questions Test B}} & \rotatebox{90}{\scriptsize{Wiki Trivia Test A}} &
   \rotatebox{90}{\scriptsize{Wiki Trivia Test B}} \\  
   \midrule
   {\scriptsize BM25} & \multirow{4}{*}{\rotatebox{90}{{\scriptsize NDCG@10}}}  & 61.00 & 58.31 & 79.14 & 81.35 & 26.04 & 24.20 \\[2pt]
   \cline{3-8}\\[-1ex]
   {\scriptsize mMiniLM}           & & 77.16 & 74.62 & \textbf{84.76} & 85.10 & 37.59 & 38.82 \\[3pt]
   {\scriptsize HerBERT}$_{base}$ & & 75.91 & 73.23 & 81.96 & 81.73 & 34.91 & 36.06 \\[3pt]
   {\scriptsize HerBERT}$_{large}$ & & 82.03 & \textbf{80.16} & 82.20 & \textbf{85.89} & \textbf{40.07} & \textbf{40.74} \\[3pt]
   \cline{3-8}\\[-1ex]
   {\scriptsize T5}$_{base}$      & & 78.39 & 74.62 & 82.94 & 84.54 & 39.67 & 38.98 \\[3pt]
   {\scriptsize T5}$_{large}$      & & \textbf{82.75} & 77.80 & 83.33 & 84.39 & 39.29 & 40.61 \\[3pt]
   \cline{3-8}\\[-1ex]
   {\scriptsize ColBERT}           & & 75.78 & 72.00 & 83.39 & 82.99 & 36.30 & 36.94 \\[3pt]
   \midrule
   {\scriptsize BM25} & \multirow{4}{*}{\rotatebox{90}{{\scriptsize MRR@10}}} & 55.01 & 52.00 & 83.56 & 86.90 & 35.14 & 33.66 \\[3pt]
   \cline{3-8}\\[-1ex]
   {\scriptsize mMiniLM}           & & 72.54 & 70.27 & \textbf{92.22} & \textbf{92.15} & 51.94 & 54.59 \\[3pt]
   {\scriptsize HerBERT}$_{base}$ & & 71.42 & 68.19 & 87.67 & 88.49 & 48.57 & 49.80 \\[3pt]
   {\scriptsize HerBERT}$_{large}$ & & 78.39 & \textbf{76.38} & 87.93 & 91.24 & \textbf{57.23} & \textbf{57.68} \\[3pt]
   \cline{3-8}\\[-1ex]
   {\scriptsize T5}$_{base}$      & & 74.84 & 68.59 & 89.24 & 91.30 & 55.86 & 55.12 \\[3pt]
   {\scriptsize T5}$_{large}$     & & \textbf{79.45} & 74.19 & 89.82 & 91.36 & 55.18 & 57.50 \\[3pt]
   \cline{3-8}\\[-1ex]
   {\scriptsize ColBERT}           & & 72.01 & 67.97 & 89.50 & 90.08 & 50.72 & 51.70 \\[3pt]
   \bottomrule
\end{tabular}
\caption{A comparison of retriever and rerankers models evaluated on data subsets from PolEval dataset. To evaluate the models, we used NDCG @10 and MRR @10 evaluation metrics.}
\label{tab:poleval}
\end{center}
\vspace{10pt}
\end{table}

Finally, we evaluated rerankers models on test datasets available in PolEval Passage Retrieval task \cite{KNOWCON20235627}.
The rerankers models improve the BM25 rankings significantly, as shown in Table~\ref{tab:poleval}. The high scores suggest that the models were able to learn from the translated version of MS Marco included in BEIR-PL and perform effectively on datasets originally annotated in the Polish language across various domains. In most tests, HerBERT-large outperformed the other models.





\section{Conclusions}


Information Retrieval in the Polish language is still developing and is an intensive research area. Therefore, there is a great need for resources to enable further training and more accurate evaluation of existing and new deep neural IR models. In this work, we introduced the translated BEIR-PL benchmark and showed the results of a broad family of IR baseline models. We would like to encourage other researchers to participate in further development of Polish and multilingual IR models using our new resource.

Our findings revealed that IR models perform differently depending on the dataset's characteristics. In some cases, lexical similarity is the right choice to solve the task, and in other cases, it is beneficial to rely on the transformer rerankers models. Looking at the results, there is a difference between sequence-to-sequence plT5 and HerBERT based rerankers, which may point other researches into the right model choice for their use. The late interaction ColBERT reranker might be a good choice when the application demands a faster reranking model that delivers strong performance across all semantic similarity tasks.

Finally, we tested models fine-tuned on the Polish MS-Marco using Polish datasets from the PolEval competition. Our results demonstrated a significant improvement over the BM25 baseline performance across three distinct domains. Therefore, the translation quality of the provided benchmark is sufficient for Information Retrieval task.

\section{Acknowledgements}
This work was supported by the European Regional Development Fund as a part of the 2014-2020 Smart Growth Operational Programme: (1) Intelligent travel search system based on natural language understanding algorithms, project no. POIR.01.01.01-00-0798/19; (2) CLARIN - Common Language Resources and Technology Infrastructure, project no. POIR.04.02.00-00C002/19

\nocite{*}
\section{Bibliographical References}\label{sec:reference}

\bibliographystyle{lrec-coling2024-natbib}
\bibliography{lrec-coling2024-example}


\appendix

\section{Evaluation metrics\label{appedix_metric}}

\begin{itemize}
    \item \emph{Mean Reciprocal Rate} (MRR@k) -- the official MS Marco metric. MRR@k measures the quality of rank regarding the first relevant passage in ranking,
    $$MRR@k = \frac{1}{|Q|}\sum_{i=1}^{|Q|}{\frac{1}{rank_{i}}}$$.
    \normalsize
    \item \emph{Normalised Cumulative Discount Gain} (nDCG@k) -- reported in the original BEIR benchmark. NCDG@k measures the quality of ranking considering all relevant passages and its position in @k retrieved documents,
    $$NDCG@k = \frac{\sum_{i=1}^{\textit{}{k(rank\_order)}}{\frac{Gain}{log_{2}(i+1)}}}{\sum_{i=1}^{\textit{}{k(real\_order)}}{\frac{Gain}{log_{2}(i+1)}}},$$ 
    \normalsize
    where Gain is equal to 1 if passage relevant and 0 otherwise.
    \item \emph{Recall} (Recall@k) cut off at the \textit{k} ranking position. The recall@k informs how many relevant documents from the collection were classified to @k ranking.
    \\ \\
    $$recall = \frac{|relevant| \cap  |retrieved|}{|relevant|}$$.
    \normalsize
    
\end{itemize}


\end{document}